\documentclass[journal]{IEEEtran}
\usepackage[final]{graphicx}
\usepackage{graphicx}
\usepackage{amsmath}
\usepackage{wrapfig}
\usepackage{multirow}
\usepackage{colortbl}
\usepackage{color}
\usepackage{soul}
\usepackage{enumerate}
\usepackage[caption=false]{subfig}
\usepackage[table]{xcolor}
\usepackage{pifont}
\usepackage{paralist}
\usepackage{siunitx}
\usepackage{siunitx}
\usepackage{flushend}
\usepackage{amssymb}
\usepackage{paralist}
\IEEEoverridecommandlockouts
\usepackage{graphicx}

\usepackage{textcomp}
\usepackage{gensymb}
\usepackage{stfloats}
\usepackage{tikz}
\usepackage{verbatim}
\usetikzlibrary{chains,positioning,shapes.symbols,arrows,shapes,shadows,trees}
\usepackage{standalone}
\usepackage{array}
\usepackage{cite}
\usepackage{cuted, nccmath}
\usepackage{lipsum}
\begin{document}
\bstctlcite{IEEEexample:BSTcontrol}
\setlength{\parskip}{0pt}

\title{Universal Transceivers: Opportunities and Future Directions for the Internet of Everything (IoE)}

\author{
Meltem Civas, Oktay Cetinkaya, Murat Kuscu, Ozgur B. Akan
      \thanks{The authors are with the Department of Electrical and Electronics Engineering, Ko\c{c} University, Istanbul, 34450, Turkey  (email: \{mcivas16, ocetinkaya, mkuscu, akan\}@ku.edu.tr).}
       \thanks{Ozgur B. Akan is also with Internet of Everything (IoE) Group, Electrical Engineering Division, Department of Engineering, University of Cambridge, Cambridge, CB3 0FA, UK (email: oba21@cam.ac.uk).}
       
\thanks{This work was supported in part by the AXA Research Fund (AXA Chair for Internet of Everything at Ko\c{c} University), The Scientific and Technological Research Council of Turkey (TUBITAK) under Grant \#120E301 and Huawei Graduate Research Scholarship.}}

\maketitle

\begin{abstract}
The Internet of Everything (IoE) is a recently introduced information and communication technology (ICT) framework promising for extending the human connectivity to the entire universe, which itself can be regarded as a natural IoE, an interconnected network of everything we perceive.  The countless number of opportunities that can be enabled by IoE through a blend of heterogeneous ICT technologies across different scales and environments and a seamless interface with the natural IoE impose several fundamental challenges, such as interoperability, ubiquitous connectivity, energy efficiency, and miniaturization. The key to address these challenges is to advance our communication technology to match the multi-scale, multi-modal, and dynamic features of the natural IoE. To this end, we introduce a new communication device concept, namely the universal IoE transceiver, that encompasses transceiver architectures that are characterized by multi-modality in communication (with modalities such as molecular, RF/THz, optical and acoustic) and in energy harvesting (with modalities such as mechanical, solar, biochemical), modularity, tunability, and scalability. Focusing on these fundamental traits, we provide an overview of the opportunities that can be opened up by micro/nanoscale universal transceiver architectures towards realizing the IoE applications. We also discuss the most pressing challenges in implementing such transceivers and briefly review the open research directions. Our discussion is particularly focused on the opportunities and challenges pertaining to the IoE physical layer, which can enable the efficient and effective design of higher-level techniques. We believe  that  such  universal  transceivers  can  pave  the  way  for seamless  connection  and  communication  with  the  universe  at a  deeper  level  and  pioneer  the  construction  of  the  forthcoming IoE  landscape. 

\end{abstract}

\section{Introduction}

The ever-growing demand for high data rates and low latency coupled with the exponential increase in the number of interconnected devices has been met by the evolution and diversification of the conventional communication technologies, particularly moving towards higher frequency bands, e.g., mm-Wave and THz bands \cite{akyildiz20206g}. Likewise, the objective of extending our control and connectivity to underexplored environments, e.g., underwater, intra-body, with the ever-increasing resolution has led to the emergence of new non-conventional communication modalities, such as acoustic communications and molecular communications (MC) \cite{akyildiz2019moving, song2019editorial}. These fast-paced developments have been further accompanied by the introduction of numerous wireless power transfer (WPT) and energy harvesting (EH) techniques that can exploit various energy sources to overcome the limitations of finite-capacity batteries and enable self-sustaining networks within the Internet of Things (IoT) framework \cite{akan2018internet, long2018energy, kamalinejad2015wireless}. Hence, the current and future landscapes of communications and networking are characterized by increasing heterogeneity of communication and power supply technologies, which are optimized only for particular scales, environments, and applications. 

The recently introduced \emph{Internet of Everything (IoE)} framework is positioned to exploit the heterogeneity of current and next-generation communication and networking technologies (both conventional and non-conventional) to extend our connectivity to the entire universe, which is itself a natural IoE, an inherently heterogeneous network of everything we perceive, whose interactions governed by the laws of physics \cite{dinc2019internet}. Maximizing the connectivity to the universal scale through the integration of different communication technologies and their close interaction with the natural IoE is expected to enable novel applications. For example, control over biological communication pathways among living entities, such as biological cells, animals, plants, through the seamless interface of our communication technologies will have broad implications for biomedical, agricultural, and environmental applications \cite{akyildiz2015internet}. As a particular example, interfacing future molecular nanosensor networks located inside the human brain to the low-latency and high-rate electromagnetic 5G wireless networks can bring the Internet of Senses to reality, enabling the nonverbal, i.e., conceptual, communication of human-body senses between individuals. The countless number of applications that can be enabled by the blend of heterogeneous technologies across different scales and environments, however, impose several fundamental challenges, such as interoperability, ubiquitous connectivity, energy efficiency, and miniaturization.

As in all networking technologies, the opportunities and limitations pertaining to the physical layer of the IoE will largely determine what we can achieve with higher-layer techniques and protocols in terms of tackling its core challenges. With this consideration, we set out to investigate the physical architecture of a universal transceiver that, we believe, can enable the efficient and effective design of higher-level techniques for the IoE. Here we coin the \emph{universal IoE transceiver} as an umbrella term, which encompasses all transceiver devices that are characterized by the key IoE attributes, such as \emph{multi-modality}, \emph{modularity}, \emph{tunability}, and \emph{scalability}, and can be embedded into any entity or device to transform them into \emph{IoE devices}. Although combinations of these traits have already been considered in the literature for various communication devices and networks, those studies are limited to particular communication technologies and applications, and thus, away from providing general insights into transceiver architectures that can address the IoE challenges. For example, multi-modal communications has been extensively studied in the context of underwater networks to optimize the connectivity by adaptively combining acoustic communications with optical or RF communications \cite{basagni2019marlin, zhao2020energy, campagnaro2018optimal}. Smart gateway architectures, providing modularity and multi-modality in terms of communication technologies and protocols, have been proposed for IoT to shield the heterogeneity of the IoT sensing layer in its interaction with the network layer \cite{guoqiang2013design}. Cognitive radio, and in general, software-defined radio technologies consist in the dynamic tunability of the transmission and reception parameters of network nodes, which gives their flexibility in overcoming the limited spectrum challenges \cite{zalonis2005flexible, arslan2007cognitive, cetinkaya2020cognitive}. Multi-modal, bio-cyber, and macro-nano interface architectures have been proposed for MC, which transduce MC signals to different signal forms including RF and optical \cite{kuscu2015modeling, koucheryavy2021review, nakano2014externally}. 

Focusing on the fundamental traits introduced above, this paper provides an overview of the opportunities that are made possible by such universal transceiver architectures. That is performed in the context of IoE, such that the overview is not focused on particular technologies or applications. However, example scenarios of practical applications accompany the discussion to justify the need for universal transceivers combining these features. We also discuss the most important challenges in implementing such transceivers and briefly review potential research directions to practically address them. Although our discussion is focused on physical layer challenges, we occasionally extend it to higher layers to provide a broader perspective. 

\section{Universal IoE Transceiver}
We envision a universal IoE transceiver as a physical device enabling the design of higher-level communication techniques and protocols that would help overcome the fundamental IoE challenges, such as interoperability, ubiquitous connectivity, energy efficiency, miniaturization, and scarcity of bandwidth. Although this paper is not proposing a particular architecture, the following set of properties delineates the universal IoE transceiver, which is investigated in this paper in terms of opportunities and challenges: 

\begin{itemize}
	\item \textbf{Multi-modality in communications:} The universal IoE transceiver is capable of transceiving communication signals in multiple forms, e.g., RF, optical, molecular, and acoustic. Supporting multiple combinations of communication modalities, the universal transceiver will bring important functionalities, such as adaptivity, multiplexing, and seamless interfacing between different modalities as discussed in Section \ref{sec:opportunities}, which will help overcome the interoperability and ubiquitous connectivity challenges of the IoE.
	\item \textbf{Multi-modality in EH:} The considered device has an architecture that can harvest energy from multiple energy sources, such as RF, mechanical, biochemical, and solar energy sources, as well as, utilize WPT in an hybrid manner. The adaptivity enabled by multi-modality in energy supply will effectively address the energy scarcity challenges in the IoE. 	
	\item \textbf{Modularity:} The device has a modular architecture that enables reconfigurability. This will allow the utilization of the most appropriate communication and EH modalities for particular applications and application environments, eliminating any redundancy. Future communication and EH technologies can also be integrated into the universal transceiver thanks to its modular structure.  
	\item \textbf{Tunability:} The dynamic tunability of transmission and reception parameters for individual communication modalities, e.g., the frequency of transmit signals in RF communications or the type of carrier molecules in MC, will contribute to the adaptivity, energy efficiency, and environment-compatibility of the universal transceiver, as discussed in Section \ref{sec:opportunities}. 
	\item \textbf{Scalability:} The device architecture is scalable to micro/nanoscales, supporting ubiquitous connectivity especially in bio-nano environments and enabling the deployment of more IoE functionalities in smaller volumes.
	\end{itemize}

In the following, we address the \textbf{PHY/device} opportunities and challenges for a universal IoE transceiver defined by aforementioned properties. Extension of the investigation to upper layers (e.g., network layer) within the IoE framework can be considered as a future work.

\section{Opportunities of Universal IoE Transceivers}
\label{sec:opportunities}

The universal IoE transceivers can present unprecedented opportunities within the emerging IoE framework by supporting the interoperability of heterogeneous communication networks through seamlessly interfacing their physical layers, providing ubiquitous operation and adaptivity at diverse and dynamic environments, and enabling the opportunistic utilization of multiple communication and energy modalities. In this section, we present a detailed discussion of these opportunities along with the potential applications.     

\subsection{Heterogeneous IoE}

The current landscape of networking is already characterized by heterogeneity of communication technologies. On top of that, our perennial quest to reach out to underexplored environments, such as underwater and intra-body, is pushing the diversification of the communication technologies to new levels. Besides, natural communication systems, which are optimized through billions of years of evolution, are also characterized by substantial heterogeneity, for example, in terms of biological signalling pathways and signalling entities. 

IoE, which aims at integrating these conventional and unconventional communication mechanisms involved in these heterogeneous networks to merge natural and human-made networks and maximize our connectivity to the universal scale, poses a great interoperability challenge. 

Nature has developed its unique and elegant ways of overcoming the interoperability challenge in interconnecting different species through inter-kingdom communication pathways which lend itself to intriguing consequences that are yet to be fully uncovered. For example, the signalling through gut-brain axis connects the nerve cells in human central nervous system and the microbes in the gastrointestinal tract, which have implications in many human-body disorders and diseases, such as depression and irritable bowel syndrome \cite{foster2013gut}.  Plants and animals communicate with diverse signalling mechanisms, e.g., visual and biochemical signals released by the plants attract the animals that can disperse the seeds of the plant, or, bat-pollinated plants can communicate with the bats via acoustic signals \cite{schoner2016acoustic}. Similarly, the plants and the microbes located at the plant's root, e.g., fungi and bacteria, communicate by means of chemical signals enabling mutual relationship for the plant and the microbes. That is, the plant releases metabolites to the soil, and the microbes provide nutrition and  protection against pathogens and stress for the plant upon stimulation \cite{harkes2020shifts}.
These inter-kingdom communication pathways are typically realized through the sharing of a common information carrier that is recognized by all involved species. 

\begin{figure}[t]
	\centering
	\includegraphics[width=0.99\columnwidth]{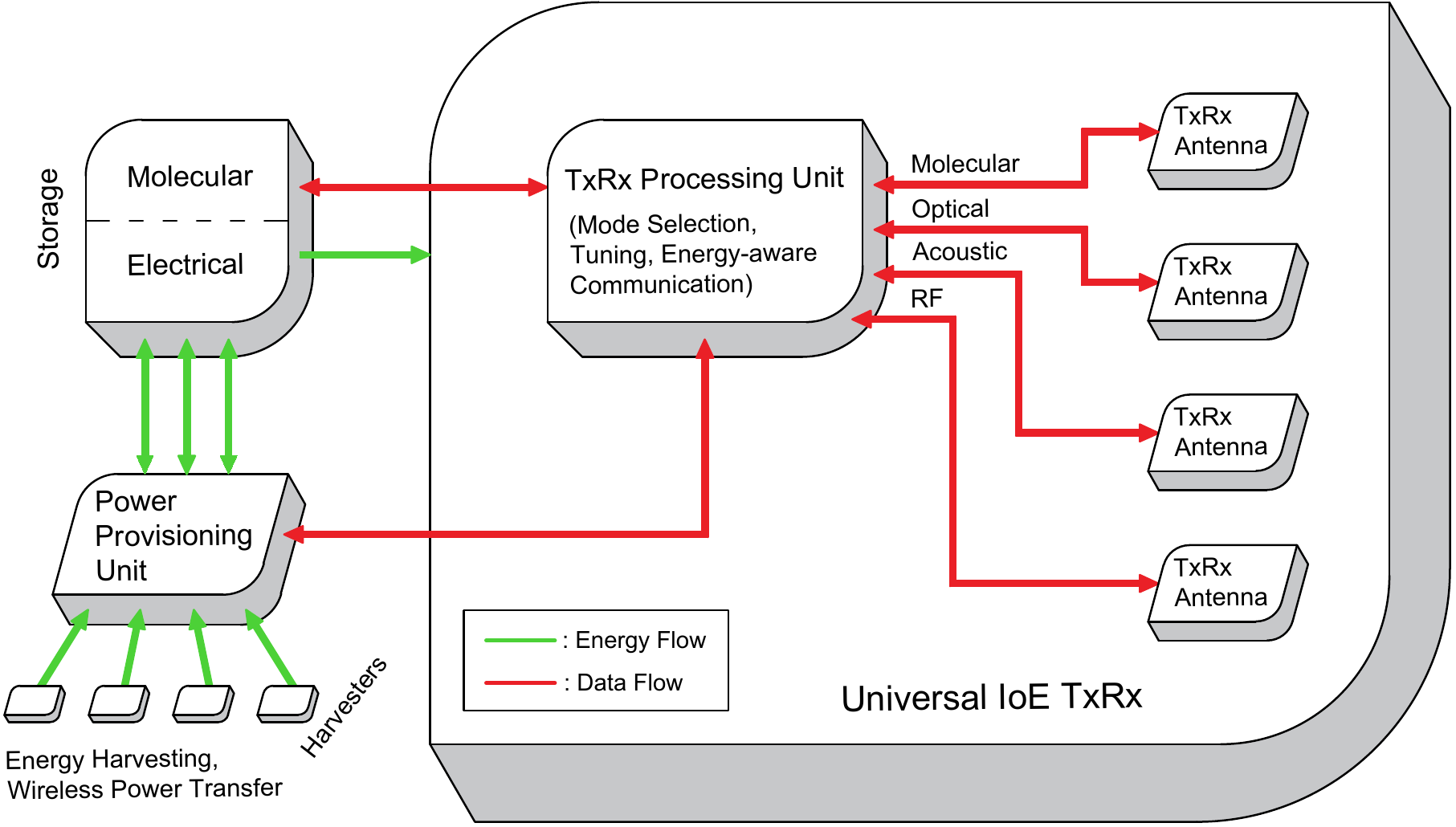}
	\caption{An example architecture for a universal IoE transceiver.}
	\label{fig:architecture}
\end{figure}

It is difficult to envision a scenario where a signal type of carriers is used to interconnect all entities, both natural and human-made, involved in a universal IoE. However, a universal transceiver supporting multi-modal communications, as shown in Fig. \ref{fig:architecture}, can overcome the interoperability challenge of the IoE by acting as a seamless interface between different types of networks.  

\begin{figure}[t]
	\centering
	\includegraphics[width=0.70\columnwidth]{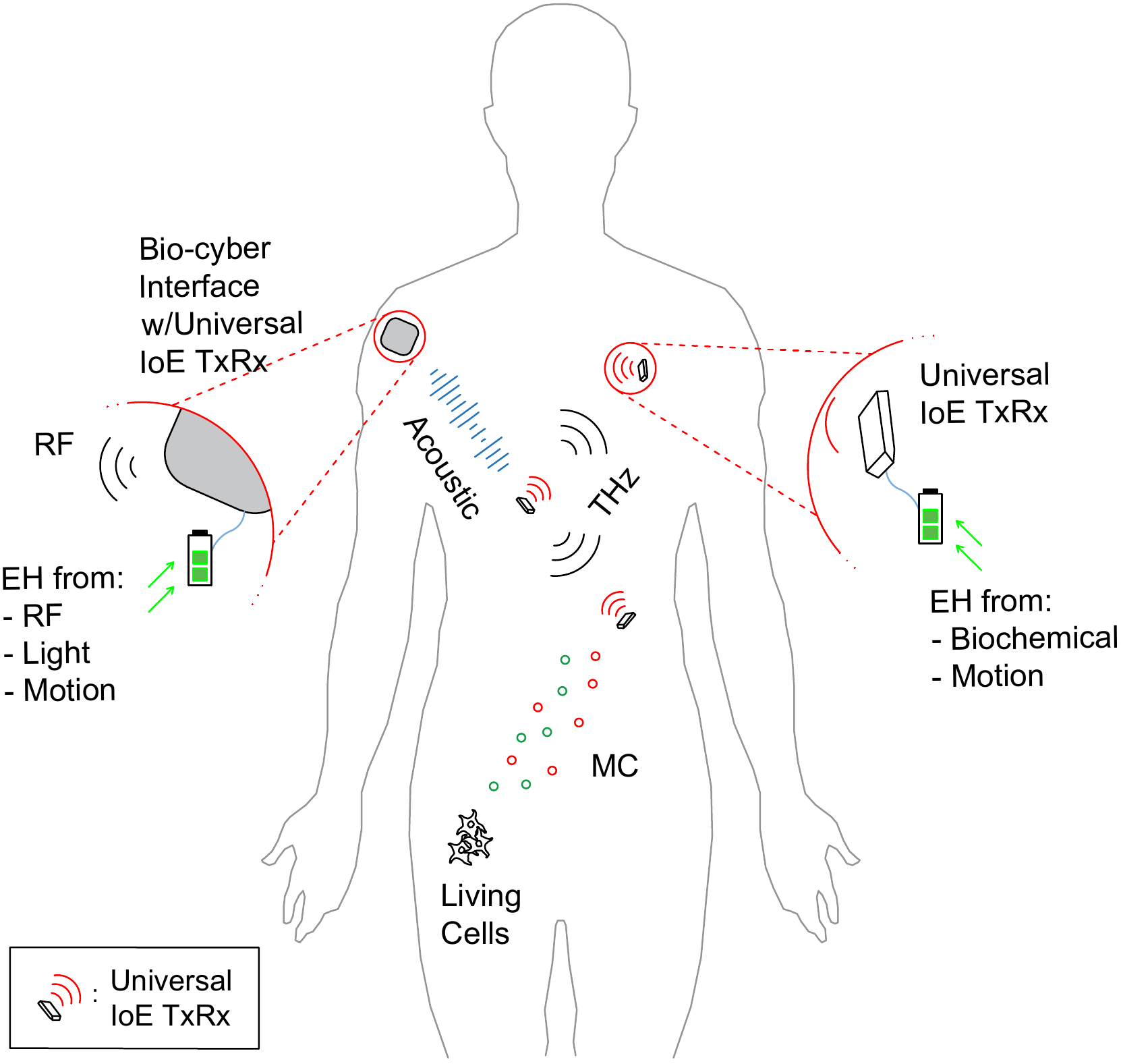}
	\caption{Remote healthcare application with continuous health monitoring enabled by universal IoE transceiver.} 
	\label{fig:IoBNT}
\end{figure}

Interfacing various networks of heterogeneous technologies via universal transceivers, that is the underlying theme of the IoE framework, is expected to open up unprecedented opportunities. For example, reprogramming biological communication systems for smart farming, can be realized by bridging different IoE technologies. More specifically, an Internet of Bio-Nano Things (IoBNT) \cite{akyildiz2015internet}, which uses molecular communication modality to regulate the growth of certain microorganisms in soil that provide the plants with the immune response against the upcoming environmental changes, can be interconnected via an electromagnetic communication modality with an Internet of Drones (IoD) \cite{long2018energy,cetinkaya2020efficient,cetinkaya2020internet} that provides real-time feedback by monitoring the crops having disease or stress. Similarly, an IoBNT using MC to communicate with the intra-body cells can be interconnected with a bio-cyber interface via multiple communication technologies suitable for the intra-body communication,  e.g., MC, THz, and acoustic, for healthcare applications such as remote healthcare with continuous monitoring as illustrated in Fig. \ref{fig:IoBNT}. In particular, an IoBNT can monitor the concentrations of hormones or molecules in the blood, and transmit this information to the bio-cyber interface, which can connect to the conventional EM networks using RF modality so that the healthcare provider can remotely monitor the person's health condition \cite{khan2020nanosensor}. Moreover, the neural interfaces and wearable sensor/actuator networks embedded with universal transceivers can be utilized in the transfer of senses, thoughts and skills non-verbally among individuals, i.e, Internet of People and Senses (IoPS) \cite{grau2014conscious}. In particular, a neural interface communicating with the neurons in the human brain can be interconnected to the other neural interfaces for the transfer of thoughts via low-latency and high-data-rate communication technologies, e.g., THz and optical, that can satisfy the requirements of such applications \cite{akan2021information}. 
The universal IoE transceivers can facilitate the cooperation of many other \emph{Internet of X's (IoXs)}, such as Internet of Vehicles (IoV) and Internet of Energy (IoEn), within the IoE framework for novel applications emerging from the close interaction between the heterogeneous technologies inherent in these IoXs.

\subsection{Ubiquitous Connectivity and Continuous Operation}
Varying environmental factors and dynamically changing application requirements demand flexibility and adaptation for maintaining reliable communication and continuous energy harvesting concurrently. The modular architecture and tunability, on which a universal IoE transceiver relies, can satisfy these requirements by enabling reconfigurability and adaptation. That is, a universal IoE transceiver either reconfigures its communication and/or energy harvesting modes to the one that is best suited for the environment and the application or opts for the most appropriate transmission/reception and energy harvesting parameters. This can provide seamless coverage and uninterrupted operation everywhere.

There exist a number of communication technologies, each of which has various advantages and disadvantages for a particular application or an environment (see Table \ref{table:comparison}). The multi-modality support of an IoE device for communication can eliminate the shortcomings of a particular technology by allowing the utilization of advantage of another in a flexible manner. As an example, consider free space optical (FSO) and RF communications for outdoor applications. FSO communication can provide high data rates compared to RF communication. However, FSO communication is highly affected by the weather conditions such as fog and clouds, and atmospheric turbulence \cite{kaushal2015free}. This can cause link failures or range reduction due to strong attenuation in hard weather conditions. To utilize the potentials of both technologies in the fullest extent and ensure reliability together, the IoE transceiver can employ both modalities adaptively and collaboratively using novel methods such as hybrid channel coding  \cite{eslami2010hybrid}.

\begin{table}[!t]
	\begin{center}
		
		\centering
		\caption{Comparison of some of the key IoE communication modalities.}
		\setlength{\extrarowheight}{1pt}
		\label{
		}
		
		\begin{varwidth}{\textwidth}
			{\scriptsize
				\begin{adjustbox}{width=0.48\textwidth}
					\small
					\centering
					\begin{tabular}{|c||c|c|c|}
						
						\cline{1-4}
						
						\textbf{Communication modality} & \textbf{Medium} & \textbf{Data rate} & \textbf{Communication range}  \\ \cline{1-4} 
						\multicolumn{1}{c}{}  \vspace{-2.2mm} \\ \hline 
						
						\multicolumn{1}{|c||}{\textit{\textbf{Molecular}}} & \begin{tabular}[c]{@{}c@{}}Intrabody,\\ Airborne,\\ Underwater\end{tabular}   & Very low
						& Low \\ \hline
						
						\multicolumn{1}{|c||}{\textbf{THz\textit{-band EM}}} & \begin{tabular}[c]{@{}c@{}}Intrabody,\\ Airborne\end{tabular}     & Very high &  Low   \\ \hline
						
						\multicolumn{1}{|c||}{\textit{\textbf{Acoustic}}}  & \begin{tabular}[c]{@{}c@{}}Intrabody,\\ Airborne,\\ Underwater\end{tabular}  & High & Moderate  \\ \hline
						
						\multicolumn{1}{|c||}{\textit{\textbf{RF}}}  &  \begin{tabular}[c]{@{}c@{}}Airborne,\\ Underwater\end{tabular} & High & High\\ \hline
						
						\multicolumn{1}{|c||}{\textit{\textbf{Optical}}}  &  \begin{tabular}[c]{@{}c@{}}Underwater,\\ Intrabody\end{tabular} & Very high & Moderate\\ \hline

					\end{tabular}
					
			\end{adjustbox}}
		\end{varwidth}
		\label{table:comparison}
		
	\end{center}
\end{table}

\begin{figure}[t]
	\centering
	\includegraphics[width=0.99\columnwidth]{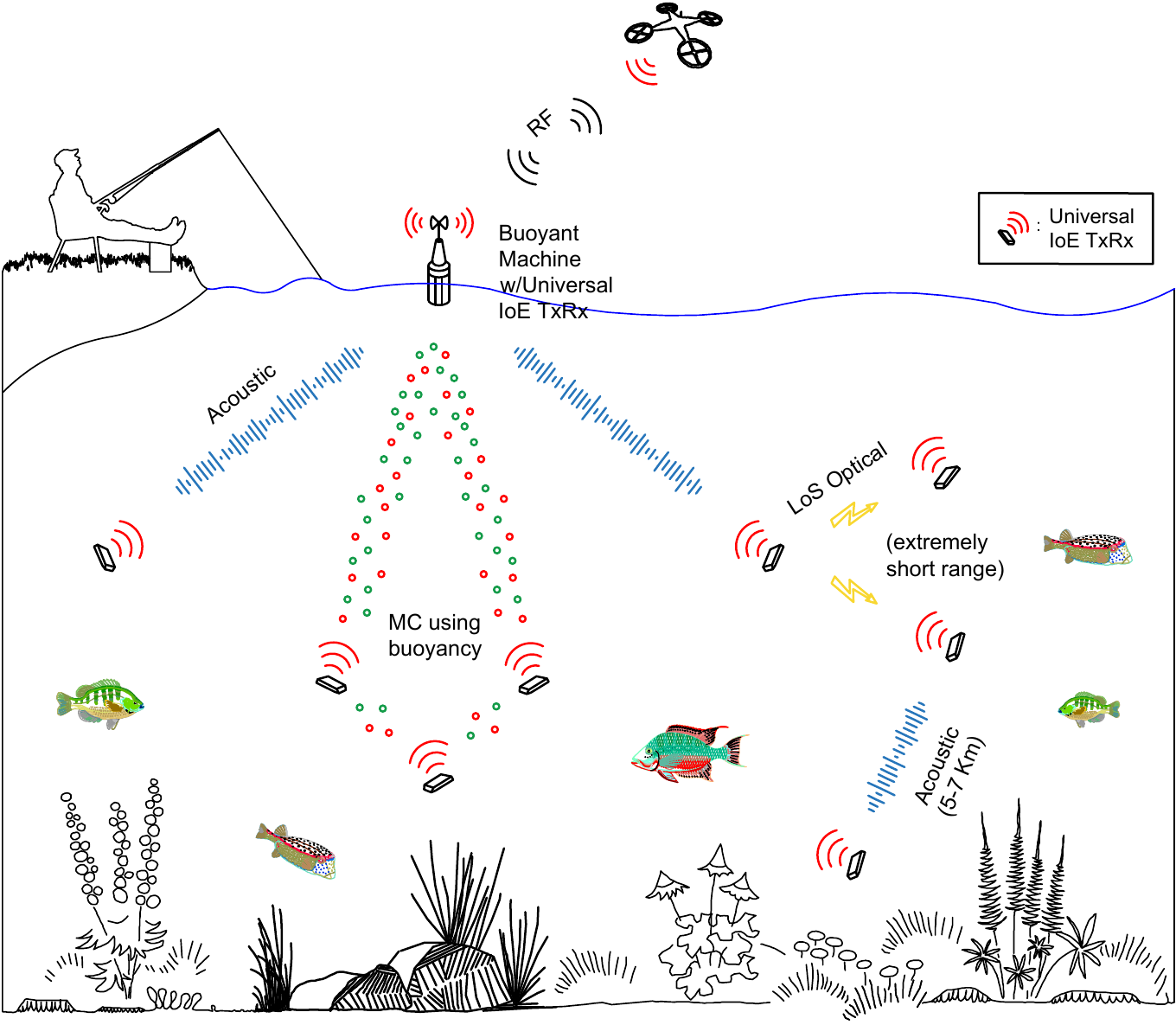}
	\caption{Ubiquitous connectivity  enabled by the universal IoE transceivers for the underwater applications. }
	\label{fig:underwater}
\end{figure}

Similarly, a universal transceiver can incorporate the advantages of multiple communication technologies, e.g., acoustic, optical and MC using buoyancy, for continuous operation in the underwater. Acoustic communication is a well known solution for underwater applications such as underwater habitat monitoring and submarine communication due to its robustness and relatively long-range support in the underwater. However, acoustic communication cannot support the applications requiring high-data rates such as high-quality video streaming due to the limited bandwidth. On the contrary, optical communication can provide extremely high data rates but suffers from the strong underwater attenuation limiting the communication range. Moreover, the optical links may not be always available because they require line-of-sight. For instance, in an underwater surveillance scenario based on the sensor network equipped with the universal IoE transceivers, videos or images can be streamed in real-time using optical links, and when the optical links are not available videos and images can be streamed using acoustic links nevertheless with low quality. Furthermore, both optical and acoustic communication may not work 
for the deep sea applications such as transmission of the location of \textit{black box} on deep ocean floor,
since optical and acoustic communication suffers from high attenuation and long channel multipath limiting the reliable communication in long distances, respectively. 
In such a scenario, the black box with a universal transceiver can use MC via buoyancy, which is a recently introduced delay-tolerant method for the deep sea applications \cite{guo2020vertical}, to transmit its location to the surface buoyant machine vertically. In MC via buoyancy, the injected information-carrying liquid, which is lighter than the ambient, is propelled by the buoyancy forces after losing the initial inertia force \cite{guo2020vertical}. Thus, when the buoyancy force is stronger than the horizontal ocean currents, buoyancy-based MC links are suitable. 
On the other hand, MC via buoyancy still requires effective coding methods to improve the low transmission rates. 
In addition, according to the availability of the sensor network with the universal transceivers deployed nearby, the universal transceiver of the black box can opportunistically utilize acoustic or optical communication to communicate with the sensor nodes that can relay the information to the surface buoyant machine. Then, the surface buoyant machine relays the location of the black box to the drone network that can provide the rescue team with knowledge of the disaster area, as illustrated in Fig. \ref{fig:underwater}.

The dynamic tunability of transmission/reception parameters for a particular modality in the universal transceiver can help overcome challenges imposed by the environmental factors, affecting reliable operation. This can be exemplified by the case of MC, which uses molecules as the information carriers. The environmental factors,   
such as the chemical composition of the channel, dynamic flow conditions, and temperature, can affect the lifetime of the information carrying molecules and the bio-compatibility of the MC link \cite{kuscu2019transmitter}. For example, considering an intrabody IoE application, a particular type of information carrying molecules might be safe only up to a certain level of concentration in blood, beyond which it may cause toxic effects for tissues and organs. A universal IoE transceiver can adapt to this condition by alternating the type of the carrier molecules within the set of molecule types harvested and stored in its molecular storage unit, as shown in Fig. \ref{fig:architecture}.

Continuous operation also requires exploitation of multiple energy sources due to the uncertainty inherent to these energy sources. A multi-modal energy harvesting unit sensing and exploiting diverse forms of energy resources available in the environment can alleviate this problem. The multi-modal energy harvesting capability will not only provide continuous operation but also increase the energy harvesting capacity of the universal IoE transceivers as discussed in Section \ref{EH_cap}.

\subsection{Increasing the Communication Performance}

Utilizing multi-modal communications for performance improvement is a emerging research topic \cite{famaey2018flexible, basu2021energy}. 

The current approaches are based on either using a particular modality at a time or using two different communication modalities simultaneously via multiplexing. 
However, with only these functionalities, the adaptation to the diverse environments is not possible. 
A universal IoE architecture can simultaneously provide a combination of diverse functionalities 
including the aforementioned ones such as mode-switching, multi-modality
and dynamic tuning of transmission/reception parameters, which can allow fullest adaptivity to the environment and/or application, and provide improved performance at the same time.

The functionalities of a universal IoE transceiver can pave the way for novel solutions tackling the performance impediments even in the arguably most challenging environments. For instance, there exist only a few technologies that can potentially support in-body communications, e.g., MC and Terahertz (THz)-band communications, due to the challenges imposed by the size constraints and physiological conditions. Considering an MC communication scenario involving multiple nanomachines sharing the same channel medium, multi-user interference (MUI) caused by the interfering molecules released by the other transmitters can degrade the performance significantly \cite{dinc2017theoretical}. In case of a crowded MC network, the universal IoE transceiver can tune or reconfigure its communication mode opportunistically to minimize or avoid interference. For instance, the IoE transceiver can alternate the type of carrier molecules prior to transmission by dynamically sensing the channel in terms of concentration levels, similar to the cognitive radio methods in EM communications \cite{kuscu2019channel}. 
Alternatively, it can utilize an orthogonal communication modality, e.g., THz-band EM communications, to avoid MUI depending on the channel conditions. In addition, hybrid communication schemes multiplexing MC and THz-band communications can be devised to exploit orthogonality of these modalities collaboratively.

The functionalities of the universal IoE transceivers can also be leveraged in transmission control. For instance, in the case of MC, synchronization between the transmitter and the receiver is challenging because the feedback link from the receiver can cause huge overhead due to slow propagation of molecules \cite{lin2016clock}. An IoE transceiver can employ different types of molecules for data transmission and synchronization \cite{jamali2017symbol}.
 Similarly, an IoE transceiver can use a communication modality in transmission control to assist another modality used in data transmission. 
A relevant example is the hybrid underwater optical/acoustic networks for real-time video streaming \cite{han2014real}. 
The optical data transmission is only possible when the optical modems are aligned properly. Therefore, until the alignment, while the optical links are not available, acoustic links can provide backup for transmission of data, and control signals, such as acknowledgements, for the optical modems.

\subsection{Increasing the Energy Harvesting Capacity}
\label{EH_cap}

EH enables IoE transceivers to operate without requiring any external source of power. By assuring theoretically unlimited energy, EH aims to increase longevity and hence to replace batteries, in particular, which are not only capacity-limited but also maintenance-heavy and environmentally unfriendly. Despite its benefits, EH often fails to support periodic reporting due to the uncontrollable and unpredictable nature of the exploitable sources and the power-hungry communication capabilities of the IoE transceivers. Hence, the coexistence of different harvesting mechanisms is crucial to tackle the non-deterministic behavior of the EH, achieving reliability in IoE communications. 

\textit{Hybrid} EH enables us to utilize ambient sources opportunistically and thus reduce the variance in the EH output thanks to the enhanced continuity in the energy input. If the deployment environment has multiple sources available together, the IoE transceiver can harvest them simultaneously or alternately. That often increases the amount of collected energy for a given time period, which expands the feasible energy tunnel enhancing the optimal energy allocation policy \cite{akan2018internet}. 
If a large enough storage is employed, the optimal energy allocation policy of the hybrid EH IoE transceiver approximates to that of an unlimited battery, referring to perpetual and reliable operations at the highest transmission rate possible.

There are a plethora of studies combining different energy sources/harvesters in practice \cite{gupta2017broadband}. The prominent examples of this exploit either a single source (providing multiple energy streams, such as temperature gradients and vibrations generated on a pipe due to water flow) with different harvesting mechanisms, e.g., thermoelectric and piezoelectric generators, or multiple (diverse) sources, such as solar and wind power or acoustic waves and RF signals. The hybridization of sources should also increase the overall energy conversion efficiency of the system, compared to the sum of individual efficiencies, which can be achieved at the slight expanse of volume, cost, and complexity \cite{seol2016hybrid}. Increasing the EH capacity is particularly important for applications in which the power provisioning process is intertwined with data communication. One example is the nano-devices harvesting molecules for energy generation, where some of those molecules are then used to transmit data. Hence, the IoE transceivers with multi-modal EH capabilities will increase the likelihood of energy generation from their surroundings by accommodating the necessary components, which offers numerous benefits especially for the multi-modal communications in the IoE domain.

\section{Challenges and Potential Research Directions}

The fundamental properties of the universal IoE transceivers, i.e., multi-modality in communications and energy harvesting, modularity, tunability and scalability, bring many opportunities, some of which are discussed in the previous section; however, the realization of them poses unique challenges. In this section, some of these challenges along with the open research directions that can potentially address them are discussed. We first focus on the open problems in enabling the multi-modality in  communications and energy harvesting. Our discussion is then extended to the adaptivity challenges in connection with the tunability property of the universal transceivers with regard to several aspects including channel sensing and energy awareness. We also provide an in-depth discussion on the miniaturization challenges to satisfy the scalability criterion of the universal transceivers. 
\subsection{Multi-modal Communications}
\label{multimodal}
Integration of multiple communication modalities is a key feature of the universal IoE transceiver architectures, providing the IoE devices with the capability of seamless interfacing between heterogeneous networks and ubiquitous operation. 
On the other hand, multi-modality in communication poses two fundamental challenges, namely, reliable and information-efficient inter-modality transduction,
and cross-modality interference. 

Inter-modality transduction is highly common in conventional communication networks. For instance, conversion between electrical and optical signals is the enabler of the global Internet through optical fiber links. Acoustic transceiver components such as acoustic signal generators and receivers also rely on transduction between acoustic and electrical signals.  On the other hand, with the emergence of IoBNT, research efforts are increasingly focused on bridging conventional modalities, e.g., optical, magnetic and electrical, with the unconventional carriers of information such as molecules \cite{liu2017using,grebenstein2018biological,vanarsdale2020redox}. In this direction, proposed transduction methods are mainly based on synthetic biology products and tools, such as whole cell biosensors and optogenetic methods \cite{dixon2021sensing}. However, entirely biological nature of these transduction methods limits their application range to in vivo environments. On the other hand, the universal IoE transceiver consisting in the interface of conventional and unconventional communication modalities must be able to operate in various environments to support the universality of the IoE. To this end, investigation of novel nanomaterials which are compatible with a large range of environments and scales, and support transduction of different energy forms into each other could be a promising approach. Transduction performances of such materials in terms of information efficiency and delay also an issue requiring through investigation along with the hardware complexity burden that each modality adds.  
Moreover, multi-modality in communications comes with its unique challenges. For example, it is imperative to investigate cross-modality interference, which could significantly affect the communication performance, if two modalities are interlinked indirectly through their impact on communication channel or transceiver. In the following, we discuss some of these challenges and propose potential research directions.

\subsubsection{Transduction}
Inter-modality transduction can be performed in two ways, namely transduction of all modalities to and from electrical domain, and direct transduction between different modalities without conversion to electrical domain. We envision that inter-modality transduction in universal IoE transceiver architectures is based on the first one because current signal processing elements rely on electrical signals providing fast processing.  Important challenges regarding the inter-modality transduction are transduction efficiency, processing delay, throughput mismatch between different modalities, and hardware complexity. Direct inter-modality transduction without conversion to electrical signal is also a potential approach that can be taken to address the aforementioned issues with the indirect transduction \cite{huang2016chemical, hajizadegan2017graphene}.

Inter-modality transduction must be information efficient such that information loss during the transduction process should be minimized. In this direction, investigation of high-performance novel materials enabling transduction from a set of diverse modalities to electrical domain and from electrical domain to the modality of choice is crucial. Employing the same material for transduction of all modalities is also crucial for integrability. Moreover, for an efficient transduction, respective material parameters, which affect the material properties, must be optimized as well. Graphene and related materials (GRMs) are promising in this direction.   Graphene is a two-dimensional carbon-based nanomaterial having unique electrical, optical, mechanical, thermal and chemical properties \cite{ferrari2015science}, which make it suitable for transduction of diverse modalities including acoustic, optical, RF/microwave, THz and molecular. The outstanding properties of graphene also enable sensitive and fast sensing of diverse modalities including light and molecules \cite{ferrari2015science}. Moreover, as a planar material graphene provides flexibility in a way that high integrability and compactness are possible \cite{ferrari2015science}.

\paragraph{GRMs in acoustic-electrical transduction}
Thanks to its high mechanical strength and ultra-thin structure, graphene is deemed promising as a wideband ultrasonic transducer as well as wide-frequency-response electrostatic acoustic transducer \cite{zhou2015graphene,zhou2013electrostatic}. 
Thin structure of graphene-based diaphragm of acoustic transducers provides the broadening in the frequency spectrum of the generated or detected acoustic waves considering conventional diaphragms, and high mechanical strength of graphene makes graphene-based diaphragms robust to deformation \cite{zhou2013electrostatic}. 
On the other hand, such acoustic transducers can have limitations in terms of the output quality. For instance, graphene-based acoustic wave generators provide limited output pressure density due to the loss caused by high sheet resistance of graphene \cite{xu2013flexible}. In this direction, thermo-acoustic graphene acoustic wave generators using thermal variations in the air near its surface instead of mechanical vibrations to generate acoustic waves offer a great potential because of their flatter output response at wide frequency range and the output power increasing linearly with the input power \cite{tu2019novel}. In this direction, investigating the performance of graphene acoustic transducers in diverse environments, e.g., underwater, is also crucial for the IoE support.    

\paragraph{GRMs in EM-electrical transduction}
Graphene can also be used in wideband optical to electrical and electrical to optical conversion thanks to its electro-absorption and electro-refraction properties \cite{romagnoli2018graphene}. Due to its high conductivity, graphene can also be utilized in radiating and receiving RF and microwave signals. Although exfoliated and chemical vapor deposition (CVD)-based graphene sheets exhibit large surface resistance causing large losses at RF/microwave frequencies \cite{gomez2012microwave}, it has been demonstrated that screen-printed graphene components can be utilized in RF sensing and transmission \cite{huang2016graphene}. 
Graphene also supports the tunable propagation of \textit{Surface Plasmon Polariton (SPP) waves} at THz band \cite{liu2014coherent}, which are EM waves propagating along a metal-dielectric (or semiconductor-dielectric) interface formed through the interaction of EM field with the oscillation of surface charges. This capability of graphene has opened the way for the development of tunable graphene plasmonic modulators modulating SPP waves, and tunable graphene plasmonic nano-antennas converting SPP waves to EM waves and vice versa \cite{jornet2014graphene,singh2016graphene}. 

\paragraph{GRMs in molecular-electrical transduction}
Graphene can also be utilized in molecular-to-electrical and electrical-to-molecular transduction \cite{kuscu2019transmitter}. 
Field-effect transistor-based biosensors (bioFETs) made of functionalized graphene have been shown to be promising in the transduction of biomolecular concentrations, e.g., present in blood, saliva and tissues \cite{szunerits2018graphene,tsang2019chemically}, into electrical signals.
There also exist porous graphene structures, which can be utilized as a membrane in drug reservoirs that can selectively release particular types of molecules through controllable pore sizes \cite{walker2017extrinsic,kuscu2019transmitter}. In addition to GRMs, transition metal dichalcogenides (TMDs) are also emerging 2D layered materials offering great potential for inter-modality transduction \cite{manzeli20172d}. A particular member of TMDs is Molybdenum disulfide (MoS$_2$), which has attracted great attention because it is electrically and mechanically robust.  
In this direction, various applications of MoS$_2$ have been reported including highly sensitive molecular-to-electrical conversion \cite{sarkar2014mos2} and optical-to-electrical conversion \cite{lopez2013ultrasensitive}, and electro-optic modulators \cite{li2017single,sun2016optical}. Black phosphorus (BP), which is the another class of 2D materials with the applications in optical detection and modulation, and molecule sensing \cite{zhou2017recent},
can be potentially utilized in inter-modality transduction. 

\paragraph{Cross-domain matching}
The investigation of high-performance materials is also crucial to address transduction and processing delay issue. Graphene is also promising in this direction. Thanks to its exceptionally 
high carrier mobility (up to ${\sim}10^5$cm$^2/$V$\cdot$s at room-temperature \cite{purdie2018cleaning}), graphene can transform light as well as THz signals into electrical signal with ultrafast response \cite{mueller2010graphene,mittendorff2013ultrafast}. Graphene can also enable tunable ultrahigh-speed processing components for THz-band communication such as frequency converters, mixers and modulators \cite{kovalev2021electrical}. 
Large surface specific area of graphene ($\sim 2600 \text{ m}^2/\text{g}$) provides rapid sensing of molecules as well \cite{yuan2013graphene}. It has been also demonstrated that graphene composites can be utilized in ultra-fast sensing of acoustic signals \cite{dinh2019ultrasensitive}.

Information throughput of different communication links might be different due to the diversity in employed communication methods, bringing the challenge of throughput-mismatch in interfacing heterogeneous networks. For instance, consider a IoBNT scenario where an intra-body nanonetwork employing MC interfaces with external networks via EM communication. Due to slow molecular diffusion and reaction dynamics, MC provides very low data rates compared to conventional EM communications. In such as scenario, when the external device sends high-rate data to an intra-body nanomachine with an universal IoE transceiver, the nanomachine may not relay the external input to the other intra-body nanomachines reliably because of the limited storage capacity, i.e, buffer overflow. Thus, novel transmission rate control techniques for limited storage capacity are required in this direction. 

\paragraph{Integration}
The integration of separate transduction, processing, and memory components for multiple communication modality can increase the hardware complexity of the universal IoE transceivers. In this regard, one of the potential research directions is the direct inter-modality transduction without conversion to electrical signals, which removes the necessity of processing in electrical domain, and also improves the transduction efficiency and delay. 
Alternatively, materials and methods enabling multi-functional capabilities, e.g., transduction, signal processing and memory, can be investigated towards reducing the complexity.

\subsubsection{Cross-modality Interference}
As the universal IoE transceivers can use multiple communication modalities simultaneously, a novel performance limiting factor, namely the cross-modality interference emerges. For instance, when MC and EM communication are active at the same time, MC changes molecular composition of the medium, which in turn affects the propagation of EM waves in a way that molecules vibrate as a result of energy absorption from EM waves and causes the molecular absorption noise at the corresponding EM wave frequencies \cite{yang2020comprehensive}. The impact is drastic at THz and optical frequencies. Similarly, EM radiation can affect the performance of MC because the temperature fluctuations due to molecular absorption of EM energy affect molecular dynamics \cite{singh2020radiation}. 
Moreover, the molecular composition of the medium can change the reception/transmission performance of the transceiver components. For instance, absorption of interfering molecules on the graphene THz or optical antennas can change the refractive index of the antenna surface, which in turn affects the propagation of surface plasmonic waves \cite{rodrigo2015mid}. A similar effect can be observed on graphene-oxide-based optical antennas, whose fluorescent properties are affected by the pH change in the medium due to the interfering chemical molecules \cite{gao2021graphene}.
This challenge can be addressed by identifying all the possible cross-modality interferences and developing the necessary channel and network coding methods.

\subsection{Multimodal Energy Harvesting}

Leaving aside the continuous efforts to improve simplicity and hence energy efficiency of the communication methods, such as modulation and detection techniques \cite{kuscu2019transmitter}, the most promising solution to enable self-sustaining operations is the integration of EH modalities into the IoE domain. EH has recently received tremendous research interest partly due to the energy needs set by the emerging IoE applications. Different application requirements and ever-changing medium conditions have proliferated the need for EH methodologies to achieve robust, continuous, and self-sustaining operations by reducing the impact of unpredictable and uncontrollable factors, i.e., the uncertainties. These are the options that have been considered for self-sustainable communication networks; thus, they are promising for integration into the IoE transceivers in a multi-modal manner. 
Depending on the application environment and device architectures, various energy sources can be exploited by the IoE devices \cite{jayakumar2014powering, dinc2019internet}. 
For example, solar power, flow energy, vibrations, electromagnetic signals, and metabolic sources have been deemed feasible for harvesting \cite{akan2018internet}. Despite being useful, using only one harvesting mechanism may not always be enough for the continuous operation of IoE devices. Hence, the research has looked towards \textit{hybridization} of harvesters/resources.

\subsubsection{Hybrid Energy Harvesting Methods}
An interesting research direction in parallel with the wider IoE vision is towards hybrid EH systems exploiting multiple energy sources. We have previously investigated novel EH methods based on ambient electric field \cite{cetinkaya2017electric, cetinkaya2017electric_wcom, pehlivanoglu2018harvesting}, and introduced energy-neutral Internet of Drones concept under the  umbrella of IoE \cite{long2018energy}. The aim here should be to combine several EH mechanisms for designing self-sustaining battery-less universal transceivers that can support various applications from micro-nano scale in-body applications to macro-scale airborne applications. Towards this objective, total energy consumption of universal transceiver should be estimated by considering energy requirements of different communication modalities and processing units. At the end, different EH and WPT techniques can be combined to provide the energy required for node operation, i.e. sensing and processing, as well as data communication between nodes and information gateways.

Concerning the intrabody and body area applications, human body stands as a vast source of energy in the form of mechanical vibrations resulting from body movements, respiration, heartbeat, and blood flow in vessels, thermal energy resulting from varying body temperature, and biochemical energy resulting from metabolic reactions and physiological processes \cite{dagdeviren2017energy}. Literature now includes a multitude of successful applications of EH from human body to power miniature biomedical devices and implants. Some examples are thermoelectric EH from body heat for wearable devices \cite{leonov2013thermoelectric}, piezoelectric EH from heartbeats \cite{amin2012powering} and respiratory movements \cite{zheng2014vivo} for pacemakers, as well as biochemical EH from human perspiration \cite{jia2013epidermal}. These together with EH from chemical reactions within the body, such as glucose uptake, lactate release, and pH variations \cite{dagdeviren2017energy, shi2018implantable}, can be exploited to power the IoE transceivers. Among the potential EH mechanisms, mechanical EH has attracted the most interest. Research in this field has gained momentum with the use of flexible piezoelectric nanomaterials, such as ZnO nanowires and lead zirconate titanate (PZT), in nanogenerators, enabling energy extraction from natural and artificial vibrations with frequencies ranging from very low frequencies ($<\!1$ Hz) up to several kHz \cite{wang2012nanotechnology, akyildiz2010electromagnetic}. 
A hybrid EH architecture is also proposed for IoE comprising modules for EH from light, mechanical, thermal, and EM sources \cite{akan2018internet}. 

\subsubsection{Wireless Power Transfer}

As explained previously, the universal IoE transceivers can sustain multi-modal EH capabilities by benefiting from ambient resources that are available in their deployment environment. In some cases, some of those sources can be dedicated, i.e., external sources delivering power to the IoE transceivers remotely, which refers to WPT. WPT has seen significant advances in recent years due to increasing need for replenishing the energy reservoir of IoT devices as well as wearable and implantable devices \cite{wagih2020real}. Various forms of WPT have been considered for powering medical implants \cite{agarwal2017wireless, ho2014wireless}. For example, near-field resonant inductive coupling (NRIC)-based WPT, the oldest WPT technique, has been widely-used for cochlear implants \cite{khan2020wireless, manoufali2017near}. Other techniques include near-field capacitive coupling, mid- and far-field EM-based WPT, and acoustic WPT. The near-field capacitive and inductive coupling, however, is only efficient for distances on the order of transmitting and receiving device sizes, and for the right alignment of devices. Therefore, it might not be suitable for powering micro/nanoscale IoE tranceivers \cite{khan2020wireless}. On the other hand, radiative mid- and far-field EM-based WPTs can have looser restrictions depending on the frequency of the EM waves. 

Recent research on mmWave and THz rectennas suggests the use of high-frequency EM WPT techniques for power delivery \cite{rong2018nano}. However, for intra-body applications, the higher absorption with increasing frequency and power restrictions should be taken into account. Nonetheless, simultaneous wireless information and power transfer techniques (SWIPT) utilizing THz-band have been investigated for EM nanonetworks \cite{feng2020dynamic, rong2017simultaneous}. Similar SWIPT applications have been considered for MC, where the IoE devices use the received molecules for both decoding the information and harvesting energy \cite{guo2017smiet, deng2016enabling}. There are also applications of acoustic WPT for biomedical implants using external ultrasonic devices \cite{maleki2011ultrasonically, larson2011miniature}. Although not practically implemented yet, ultrasonic EH has been also considered for powering IoE devices with piezoelectric transducers \cite{donohoe2015powering, donohoe2017nanodevice, balasubramaniam2018wireless}.

It should be noted that both EH and WPT are still at their infancy especially for nano-molecular communications in the IoE domain. New research efforts focus on reducing the power requirements of IoE transceivers by using energy efficient protocols, conservation schemes and novel topologies, rather than focusing on EH and WPT mechanisms or their joint utilization \cite{cetinkaya2019energy}. Therefore, it stands as a major open issue to utilize various EH and WPT mechanisms to create modular energy unit for the universal transceiver framework, such that different power extraction modalities can be configured depending on the application and medium of operation.

\subsection{Adaptivity and Tunability}

\subsubsection{Channel Sensing and Cognitive Radio}
Channel sensing is essential to enable adaptivity in universal transceiver architectures. The adaptive and opportunistic selection of the most proper communication modality and dynamic tuning of transmission and reception characteristics require dynamic channel sensing in multiple modalities and in a broadband manner. Broadband sensing refers to the sensing capability that can provide tunability property to the universal IoE transceivers in the maximum possible extent. For instance,  in the case of MC, sensing as many different kinds of molecules as natural systems have is required for bio-compatibility. For EM communication, sensing at wide frequency range is needed for dynamic tunability. On the other, simultaneous channel sensing in different communication modalities may impose high complexity burden. These issues necessitate the development of low-complexity, reliable and broadband channel sensing techniques for each communication modality.
Channel sensing techniques have been widely explored for RF \cite{yucek2009survey,ozger2018energy}, optical wireless \cite{gong2015temporal,liu2017sequential,arya2021spectrum} and acoustic communications \cite{baldo2008cognitive,bicen2012spectrum}, and similar investigations have just started for MC \cite{kuscu2019channel}.

For the RF modality, spectrum sensing has received great attention because it is indispensable part of cognitive radio (CR) technology alleviating spectrum scarcity problem via dynamic spectrum access. CR capability can provide the universal transceivers with the adaptivity desired in a way that the universal transceivers can reconfigure their transmission parameters according to the available communication resources in the surroundings. 
Considering the heterogeneity of networks in the IoE environment, diverse communication technologies will utilize the RF spectrum. In this respect, one particular challenge is the identification of the signal types/transceivers occupying the bands and their corresponding transmission parameters, e.g., carrier frequency, bandwidth, modulation type, transmission protocol, number of carriers \cite{eldemerdash2016signal}, which is crucial to obtain spectrum awareness. Different from the classical signal identification techniques, deep-learning based identification does not require prior knowledge about the signals such as noise statistics, demanding extensive analysis  \cite{zheng2020spectrum, tekbiyik2020spectrum}. Hence, deep learning-based identification techniques could be promising in this direction but the complexity burden should be further considered. Regarding spectrum sensing in RF spectrum, wideband and  low-complexity compressed sensing techniques are  available, thus can be utilized for the universal transceivers \cite{sun2012wideband}.

As the frequency usage is shifting towards mmWave and THz frequencies in EM communication, there is also need for ultra-wideband spectrum sensing techniques. Considering the high-sampling rate requirement of mmWave and THz band communication technologies, compressed sensing techniques exploiting the inherent sparsity of mmWave and THz signals can be utilized in universal transceivers \cite{brighente2020estimation}. Time-varying spectrum occupancy and high path losses affecting reliable sensing are the other issues relevant for channel sensing in mmWave/THz communication modalities \cite{hamdaoui2020dynamic}.

CR has also been considered for acoustic communication systems for efficient spectrum usage. 
The concerted effort in \textit{cognitive acoustic}  have been focused on underwater acoustic communications  \cite{luo2014challenges}. The challenges identified in this direction include narrowband response of the acoustic transducers, long preamble of underwater acoustic modems, i.e., guard time or cyclic prefix signal inserted to overcome interference, and highly varying underwater channels \cite{luo2014challenges}. As discussed in Section \ref{multimodal}, graphene-based acoustic transducers, which are suitable for wideband acoustic transduction \cite{zhou2015graphene}, offer a potential to address narrowband transduction issue. Moreover, further investigation considering channel conditions peculiar to other application environments different than the underwater channels is needed towards realizing cognitive acoustic for the universal transceivers.  

Similar to the CR technology, the universal transceivers can use the MC channels opportunistically by sensing the type and concentration of molecules in the channel dynamically. For instance, if the channel is crowded with a specific type of molecule, the transceiver can use a different type of molecule to avoid MUI. However, sensing becomes challenging as the number of communication entities using MC, including bio-entities using chemical signaling, in the medium increases because there might be many types of molecules in the medium. Thus, to maintain reliable communication and to minimize the possible harmful effects on surrounding bio-processes, novel MC wideband sensing techniques recognizing as many as molecules are needed.

In addition to the modality-specific issues in channel sensing, there are novel issues such as sampling rate adjustment. Sampling rate can be different for different modalities. For instance, MC does not require high-rate sampling as the wideband RF spectrum sensing requires. However, in dynamic environments higher sampling rates may be needed. Hence, the dynamic optimization of the sampling rate is an open problem.

\subsubsection{Energy-aware Communications}
The conventional way of ensuring the longevity for energy-constrained devices is to~adopt energy-efficient communication techniques since the majority of the limited energy budget is consumed during data transmission. Radio optimization, clock synchronization, low-power sleep/wake-up, energy-efficient routing, and data reduction schemes are the main enablers of lengthened operation times \cite{cetinkaya2020internet}. 

\textit{Energy-aware} communications can be regarded as a subset of those techniques yet they have some differences in practice. For example, the low-power operation can be achieved through aggressive duty-cycling and/or wake-up radios by turning the radio on (for idle-listening or transmission) and off (for power saving) periodically \cite{longman2021wake}. In similar, multi-path routing, data compression, and network coding schemes can be adopted during the system design stage to assure reliable communications on a budget. When the energy input is not constant, i.e., the device is creating its own energy via EH, however, energy-aware communication techniques come into prominence. In such cases, the devices adapt to energy availability by changing their transmit power and/or sampling rate dynamically, besides updating their data queue by dropping some packets based on freshness or priority. The recently emerged intermittent \textit{or} transient computing techniques, on the other hand, can assure application execution under scarce EH conditions through an incremental process at each power cycle based on the energy input \cite{sliper2020energy}, \cite{balsamo2020control}. 

On the molecular level, energy-aware algorithms can be used for efficient coordination and movement EH nanonetworks, e.g., engineered bacteria tracking and eliminating harmful targets in the human body \cite{islam2020energy}. Since the molecules harvested for energy generation can also transport information, the EH process directly governs data transmissions \cite{bafghi2018diffusion}. Other approaches, such as feedback control, are also promising for energy-aware MC, especially in diffusion-based information transfer \cite{musa2020lean}. 
By adopting these technologies, the IoE transceivers can achieve energy awareness in communication in concordance with the EH modality selection mechanism.

\subsubsection{EH Modality Selection}

As discussed earlier, adaptive hybridization of different energy forms is crucial for the IoE vision as it grants unique capabilities. However, the design of a hybrid EH system is not always straightforward \cite{weddell2013survey} since each harvester/storage device often needs its own interface circuitry, which translates into increased cost and complexity. 

Most of the communication modalities utilize electrical energy, which is stored in electrical storage shown in Fig.~\ref{fig:architecture}. To ensure the flow, the system needs an input power conditioning/management/provisioning circuit to interface with the EH sources, alternate between harvesters and energy stores (molecular or electrical), or combine them when necessary. This circuit should include a rectifier for converting the collected energy into a utilizable form and preventing its flow back to the harvester(s), besides a DC-to-DC converter for maximum power tracking and providing the stable output required by the transceiver antennas. The MC modality of the IoE transceiver can also use the electrical energy for internal operations, such as sensing and processing. However, the IoE transceiver harvesting molecules need additional components, such as a chemical bank with basic building components to convert the harvested molecules into molecular information \cite{guo2017smiet}, which can be attached to the molecular storage.

Once established, the IoE transceiver can harvest energy opportunistically from any source available in its vicinity, swap between (or combine) harvesters or energy stores when the primary source disappears or when it changes its application environment, and probe its proximity using different sensors enabled/disabled according to the requirements of different tasks/applications. In this way, universal, modular, and adaptive IoE systems can be devised. 

As discussed, the EH modality selection can be either pre-planned or adaptive to the ambient factors. One example is swapping sunlight to wind and/or rain drops when the weather in the deployment environment changes. The other case is the change of the environment itself, i.e., the system can harvest energy from heartbeat and/or blood flow instead of lactate or glucose when it moves within the body. These approaches often refer to \textit{online energy profiling}, in which the IoE system adapts its operation to the current channel/medium state, thereby harvesting energy opportunistically. The conventional EH transceivers, however, commonly accommodate \textit{offline energy profiling}, i.e., the medium-specific energy availability is known prior to operation. If the IoE transceiver employs the latter, it changes its application environment as it knows that a particular source will be available/in abundance in a particular location. This availability can be either due to perpetual bio-mechanical drivers, such as heartbeat, blood pressure, and breathing, or due to recurring events, e.g., periodically released hormones \cite{conn2013pulsatility}. Hence, the IoE system not only changes its medium but also its EH modality as a reflex to sustain reliable communications by getting the most out of the surrounding resources. That requires seamless coordination and efficient interaction of system components.

\subsection{Miniaturization}
The miniaturization trend at all levels of communication systems is being driven by the increasing number of connected devices. The miniaturization can provide key advantages including easy integrability, more functionality with lower footprint, increased performance, reduced power consumption and manufacturing cost. In the context of IoE, which aims at the ubiquitous connectivity of highly diverse entities, including bio/nanomachines, the miniaturization of transceiver components to micro/nanoscales is indispensable. The future 6G networks will also be a part of the IoE environment, for which the frequency shift towards THz band in EM communication is inevitable \cite{dang2020should}.  
The usage of THz band necessitates downscaling the size of transceiver components to micro/nanoscales. Moreover,  nano and bio-nano networks using unconventional communications at nanoscales, such as MC, are important components of the IoE, and interfacing these networks also requires miniaturization to micro/nanoscales.  
As such, universal transceivers must be integrable to a very large set of diverse devices and entities of varying scale and structures. Moreover, micro/nanoscale devices and components are known to be more energy-efficient in providing the similar functions as their macroscale counterparts. The miniaturization of the universal transceivers can also provide flexibility in their integration to various devices of different size scales to transform them into IoE devices. For example, micro/nanoscale universal transceivers can be embedded into both micro/nanoscale biosensors and macroscale bio-cyber interfaces as part of an IoBNT application such as remote healthcare with continuous health monitoring based on IoBNT as illustrated in Fig. \ref{fig:IoBNT}.

In the past decades, silicon semiconductor technology has been the driving force for down-scaling the size of transistors so that they can be packed onto smaller and smaller chips. However, it is now long-evident that the silicon technology is reaching its physical limits. As silicon layers are getting thinner, electron flow through silicon channels is slowing down due to the surface imperfections that are scattering charges \cite{li20192d}. New novel nanomaterials with exceptional electrical, mechanical and bio-chemical properties, such as GRMs as discussed in Section \ref{multimodal}, have emerged to overcome this limitation and to expand the diversity in functionalities in line with the More than Moore approach. 
However, there still exist substantial challenges with regard to growth and processing of these materials \cite{neumaier2019integrating}.

There also exist technical challenges to build some transceiver components with desired performances at micro/nanoscales. 
For instance, regarding conventional communications, e.g., RF and acoustic, shrinking the size of transceiver components such as antennas and modulators to micro/nanoscales while preserving their functions is still an issue because the operation in the micro/nanoscale  limits the carrier frequency to high-frequency ranges for these technologies.    
On the other hand, it should be noted that realizing universal IoE transceivers requires departure from the approaches aiming to optimize the performances of the transceiver components for individual modalities.  Rather, the underlying theme of universal transceivers must be the maximization of the versatility stemming from multi-modality, scalability, and tunability, which can nevertheless be obtained by transceiver components suboptimally performing in individual modalities. 
Accordingly, in the following, we discuss the miniaturization challenges and potential research directions in developing miniaturized IoE universal transceiver components with regard to several modalities. 

With the recent paradigm shift towards the usage of THz band, EM antennas are getting shrunk down to micro/nanoscales. Due to high losses introduced at THz frequencies by spreading, molecular absorption and scattering, the THz communication range is substantially limited. To overcome this limitation, multi-input-multi-output (MIMO) array architectures enabled by graphene plasmonic nano antennas have been proposed \cite{xu2014design,akyildiz2016realizing}. Howeover, THz wave propagation introduces novel physical phenomena that can affect the performance of MIMO arrays. For instance, a THz beam may diverge from the direction of its modulation sidebands slightly due to frequency-dependent diffraction nature of THz signals \cite{ma2017frequency}. This may degrade the performance of a MIMO array in terms of bit error rate because different antennas may detect different spectral information \cite{ma2017frequency}. Thus, novel detection and demodulation methods are needed to overcome this limitation. 
Other important challenges towards THz communications with miniaturized transceivers are the efficient signal generation and detection. In this direction, concerted effort has been dedicated to solid-state electronics and photonics approaches \cite{sengupta2018terahertz}. However, the electronics approach suffers from limited output power at THz frequencies and photonics approach based on quantum cascade lasers offers poor performance at room temperature \cite{nafari2018plasmonic}. 
In this direction, micro/nanoscale graphene transistors with extremely high carrier mobility can pave the way for the development of THz-based signal generation and sensitive detection components \cite{bianco2015terahertz}.

Silicon technology has achieved the miniaturization of optical communication components to microscale \cite{xu2005micrometre,vahala2003optical}. Recently, optical components utilizing SPPs have emerged for further miniaturization that are able to confine the light into nanometer scale. In this direction, plasmonic nanoantennas \cite{cohen2015enabling}, nanolasers \cite{lu2012plasmonic} and  modulators \cite{klein20192d} have been developed. The most pressing challenges in this direction are tuning plasmonic resonance frequency to higher frequencies, e.g., visible and near-infrared, which could be achieved through high level of doping \cite{yu20172d}, and integration of plasmonic components into non-plasmonic structures without high insertion losses \cite{haffner2015all}. 

\begin{figure}[t]
	\centering
	\includegraphics[width=0.99\columnwidth]{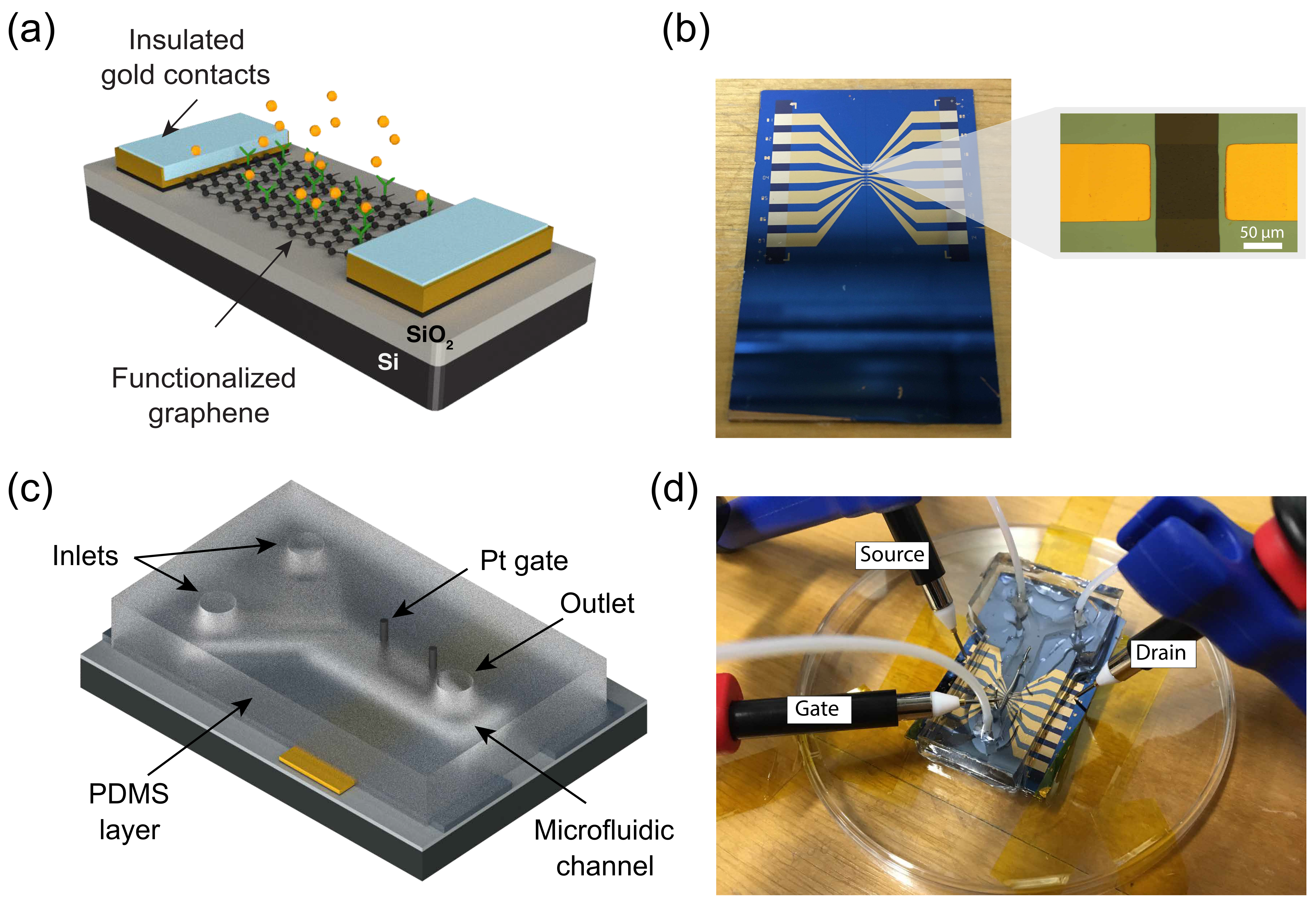}
	\caption{Graphene-based micro/nanoscale MC receiver \cite{kuscu2021graphene}: (a) Conceptual drawing of MC receiver with functionalized graphene, (b) Fabricated MC receiver, (c, d) Integration of the MC receiver into a microfluidic MC channel. } 
	\label{fig:MCreceiver}
\end{figure}

Regarding MC, the first prototype of a nanoscale MC receiver, as illustrated in Fig. \ref{fig:MCreceiver}, has been recently introduced in \cite{kuscu2021graphene}. This architecture incorporates the sensing mechanisms of electrical biosensors for detecting MC signals. 
Regarding the miniaturized MC transmitter component of the universal transceiver, as discussed in Section \ref{multimodal}, porous graphene structures enabling controllable permeability across pores via electric field  are promising for providing selective and tunable molecule transmission. On the other hand, high surface-to-volume ratio in nanopores exhibits new physical phenomenon that is not observed at larger scales. That is, surface charges become dominant in molecular transport by introducing electrostatic forces affecting the interactions between molecules and surfaces in liquids \cite{schoch2008transport}. Thus, the effect of surface charges need to be considered in MC transmitter design.  
Molecule storage is also an important challenge regarding the miniaturization of MC transmitters. In this direction, stimuli-responsive hydrogels, which are also envisioned for smart drug delivery, are promising \cite{li2016designing}. Various type of molecules can be harvested from a fluidic medium by triggering the deswelling state in a hydrogel reservoir and loaded molecules can be transferred to MC transmitter antenna, i.e., tunable porous graphene membrane, by activating the swelling state through the application of a  stimulus such as electrical field, magnetic field or light \cite{shi2019bioactuators} in a controllable manner.

The miniaturization efforts regarding RF applications have focused on carbon-based materials, especially carbon nanotube (CNT) in the last two decades. CNTs have material properties including high mobility and small size that are suitable for miniaturized RF applications such as modulators, mixers, amplifiers \cite{schroter2013carbon}. Initial demonstrations of CNTs in RF applications were nanotube radios \cite{jensen2007nanotube, rutherglen2007carbon}. Although the AM demodulator based on a single CNT in \cite{rutherglen2007carbon} has small size, i.e., \SI{1e-21}{\cubic\metre} of system volume, the overall volume of the system including the external antenna is of order \SI{1e-3}{\cubic\metre} \cite{burke2010towards}. In this direction, resonant CNT antennas can further reduce the overall system size because slow propagation of the waves on CNTs compared to free space provides a reduction in the resonant antenna size up to 100 times \cite{burke2006quantitative}.  Different from the radios in completely electrical nature, the operation of the nanotube radio implemented using a single CNT in \cite{jensen2007nanotube} is partly mechanical. Micro/nanometers long CNTs for nanoelectromechanical-based RF applications are suitable for the operation in MHz and GHz range, respectively. 
On the other hand, for both nanotube radios, the sensitivity is low such that receiving weak RF signals is not possible \cite{rutherglen2009nanotube}. Thus, further research is needed in this direction towards realizing micro/nanoscale RF components for the universal IoE transceivers. Alternatively, magneto-electric (ME) materials are also promising to miniaturize RF components such as antennas to micro/nanoscales \cite{nan2017acoustically,chen2020ultra}. For instance, 
instead of EM wave resonance, nanomechanical ME antennas operate at the acoustic resonance frequency. Since the acoustic wave length is smaller than the EM wave length at the same frequency, ME antennas can potentially provide dramatic miniaturization, i.e., hundreds to thousands of times smaller antennas, without performance degradation \cite{lin2018nems}.

Two-dimensional nanomaterial technology have also opened the way for reducing the size of acoustic transceiver components, such as
microphones, loudspeakers and earphones. In particular, graphene membranes that can generate and sense acoustic waves have received great interest due to their exceptional mechanical and electrical properties \cite{zhou2015graphene,zhou2013electrostatic,tian2015flexible}. However, downscaling the dimensions of a graphene membrane, which is a vibrating element, to micro/nanoscales limits the operation frequency range to MHz-GHz because the resonance frequency of a graphene membrane $f$ scales as $t/L^2$, where $t$ and $L$ is the thickness and the length of the graphene membrane, respectively  \cite{bunch2007electromechanical}. Thus, for the operation in the low  frequency range, i.e., Hz-kHz, large membranes in millimeter scale are required \cite{al2018tunable}. Alternatively, based on the relation $t/L^2$, thick and small scale multilayer graphene membranes can be utilized in the low frequency range \cite{todorovic2015multilayer}. However,  increased thickness may lower the sensitivity of the graphene membrane \cite{al2018tunable}. In addition, integrating graphene membrane  to the universal transceivers can be challenging since it is a mechanical component. Therefore, further research is needed in this direction to realize micro/nanoscale broadband acoustic components for the universal transceivers.

\section{Conclusion}

In this paper, we introduced the concept of universal IoE transceiver that defines the transceiver architectures, which possess multi-modality in communication and energy harvesting, modularity, tunability, and scalability. Through several practical cases involving also non-conventional communication modalities, e.g., molecular, THz-band communications, which will be relevant in the upcoming IoE landscape, we argued that such universal transceivers that can be scaled down to micro/nanoscales could address the unique IoE challenges, such as interoperability, ubiquitous connectivity, energy efficiency, and miniaturization, and could open up further opportunities in diversifying the IoE applications. Along this direction, key physical layer challenges are outlined with potential solutions, which are centered around the use of novel nanomaterials, such as graphene, that manifest unique electrical, optical, mechanical and biochemical properties.

\bibliographystyle{IEEEtran}
\bibliography{References}
\end{document}